\documentclass[epsfig,psfig]{cjpsuppl}
\begin{document}

\title{Instanton effects in high energy processes.}
\authori{A.\,E. Dorokhov}
\addressi{Bogoliubov Laboratory of Theoretical Physics,\\
Joint Institute for Nuclear Research,
141980 Dubna, Russia} \authorii{} \addressii{} \authoriii{} \addressiii{} %
\authoriv{} \addressiv{} \authorv{} \addressv{} \authorvi{} \addressvi{} %
\headtitle{Instanton effects \ldots} \headauthor{A.E. Dorokhov} %
\lastevenhead{A.E. Dorokhov: Instanton effects \ldots}

%%%%%%%%%%%%%% Pro editory supplementu: %%%%%%%%%%%%%%%
\refnum{}%slouzi editorum pro evidenci; nakonec {}
\daterec{20 October 2002;\\final version 31 December 2003} \suppl{A} \year{%
2003} \setcounter{page}{21} %\firstpage{1}
%\lastpage{000}
%\makefirsttitle
%%%%%%%%%%%%%%%%%%%%%%%%%%%%%%%%%%%%%%%%%%%%%%

\author{}
\maketitle

\begin{abstract}
Manifestation of the nonperturbative effects in hadron processes at high energy
is discussed within the instanton liquid model. Their role in high energy
diffractive quark-quark scattering, quark form factor and hard exclusive processes
with pion participation is illuminated.

\end{abstract}

\section{Introduction}

The very powerful QCD perturbative theory (pQCD) is developed in order to describe
the hadron processes at high energies. The scaling and its logarithmic violation
are described by the pQCD calculations in the lowest orders of expansion
in the strong coupling constant. Coming down in energy more and more
powers of the strong coupling constant has to be taken into account.
Moreover, in the intermediate energy region the power corrections come into play that are very sensitive to
intrinsic hadron structure.
Typically, the coefficients of the expansion in the coupling constant and
in powers of the momentum transfer are the quark-gluon matrix elements taken
at hadronic energy scale that have to
be found by nonperturbative methods. These matrix elements are governed by the evolution
equations that are determined within pQCD for different hard processes.
These equations start to be applicable at momentum transfer squared
of order 1 GeV$^{2}$ or more where the strong coupling constant becomes
small. So, it is necessary to determine the initial data for the evolution
equation which is the nonperturbative problem. Another nontrivial situation arises when at high
energy two or more hard scales appear. In that case in order to make predictions
reliable it is necessary to resum the soft part of the quark-gluon interaction to all orders.
Again the presence of the non-perturbative effects may be important in this energy region. There
are different approaches to treat manifestation of nonperturbative phenomena
at high energies: QCD sum rules, lattice QCD, quark models, {\it etc}. In this talk
we consider few examples of description of the nonperturbative effects in the hadron processes at
high energies by applying the instanton liquid model of the QCD
vacuum.

Instanton liquid model being effective model of the QCD vacuum describes
well the hadrons at low as well at intermediate energies. Thus, contact with
perturbative QCD results is possible providing the unique information about
the quark-gluon distribution functions in the QCD vacuum and hadrons at low
energy normalization point.
The various aspects of the instanton induced effects in the high energy
hadronic processes had been addressed at the very beginning of the instanton
era (see, \textit{e.g.}, \cite{Andrei:xg}). In recent works
\cite{Shuryak:2000df,Dorokhov:2002qf} the interest to them has
revived and the hope of direct detection of the QCD instanton induced
effects has appeared \cite{Moch:1996bs}. We investigate the role of instantons in diffractive
quark-quark scattering, their resummation for the quark form factor and as a
model for pion bound state describing the distribution amplitudes in hard
exclusive processes.

\section{Instanton Model of Pomeron (Landshoff-Nachtmann model)}

Soft hadronic collisions are described successfully using
Regge phenomenology, with the Pomeron exchange being dominating at high energy.
The Pomeron is considered as an effective exchange in the $t$ channel by the object
with vacuum quantum
numbers. That is why the idea that the nontrivial structure of the QCD vacuum is
relevant in describing its mechanism.

To illustrate this let us consider high energy diffractive quark-quark
scattering, where there is hope that for small momentum transfer the
nonperturbative effects give dominant contribution. One of the simplest models
of the Pomeron is based on the use of exchange by two nonperturbative
gluons. The nonperturbative part of the gluon propagator is given by (in the
Feynman gauge)
\begin{equation}
\left\langle 0\left\vert :A_{\mu }(x)A_{\nu }(0):\right\vert 0\right\rangle
=g_{\mu \nu }\int \frac{d^{4}k}{\left( 2\pi \right) ^{4}}%
e^{-ikx}D_{np}(k^{2}).
\end{equation}
In the Abelian gauge model considered originally by Landshoff and Nachtmann
\cite{Landshoff:1986yj} the nonperturbative gluon propagator $D_{np}(k^{2})$
is related to the correlation function describing the gauge invariant gluon
field strength correlator (nonlocal gluon condensate). This correlator in
general non-abelian case has form%
\begin{equation}
\left\langle 0\left\vert :G_{\mu \nu }(x)\mathcal{P}\exp \left[
ig\int_{0}^{x}dz^{\alpha }A_{\alpha }(z)\right] G_{\rho \sigma
}(0):\right\vert 0\right\rangle =
\end{equation}
\[
=\int \frac{d^{4}k}{\left( 2\pi \right) ^{4}}e^{-ikx}\left\{ \left(
D_{0}(k^{2})+D_{1}(k^{2})\right) k^{2}\left( g_{\mu \rho }g_{\nu \sigma
}-g_{\mu \sigma }g_{\nu \rho }\right) +\right.
\]%
\[
\left. +D_{1}(k^{2})\left( k_{\mu }k_{\rho }g_{\nu \sigma }-k_{\mu
}k_{\sigma }g_{\nu \rho }+k_{\nu }k_{\sigma }g_{\mu \rho }-k_{\nu }k_{\rho
}g_{\mu \sigma }\right) \right\} ,
\]%
where the first tensor structure is called non-abelian part and the second one
is abelian part. Indeed, in the abelian gauge model without monopoles $%
D_{0}(k^{2})\equiv 0$, and $D_{1}(k^{2})=D_{np}(k^{2})$. It is this property that
has been used in \cite{Landshoff:1986yj} to relate the Pomeron properties to the value of the
gluon condensate.

However, in the non-abelian model one has opposite situation.
Really, for the QCD instantons we find \cite{Dorokhov:1997iv}
$D_{1}(k^{2})\equiv 0$ and $D_{0}(k^{2})$ is nonzero.
In the realistic model of the QCD vacuum, where
the interaction with vacuum fields of large scale, $R$, is important,
the instanton cease
to be exact solution of the equations of motion, but the so called constraint instanton approximate
solution (CI) can be constructed \cite{Dorokhov:1999ig}.
This name is due to necessity to put constraints on the system to stabilize the instanton
in the external vacuum medium.
It was shown that the constraint instanton has exponentially decreasing at large
distances ($\sim R$) asymptotics.
The constraint instanton has topological number
$\pm 1$ as an instanton, however it is not self-dual field. Thus, in
realistic QCD small part of $D_{1}(k^{2})$ appears. Very similar
results have been found in the lattice simulations of the gluon field
strength correlator \cite{D'Elia:1997ne}.

Thus, within the non-abelian models there is no direct connection between the gluon propagator and the gluon field
strength correlator. So, let us directly consider the instanton part of the gluon propagator. The Fourier transform of
the instanton field is defined as
\begin{equation}
\widetilde{A}_{\mu }^{a}(p)=\eta _{\mu \nu }^{a}p_{\nu }\widetilde{\varphi }%
(p^{2}),  \label{Agen,F}
\end{equation}%
where
\begin{equation}
\widetilde{\varphi }(p^{2})=\frac{4\pi i}{p^{2}}\int_{0}^{\infty
}dss^{3}J_{2}(\left\vert p\right\vert s)\varphi (s^{2}),  \label{Phi,F}
\end{equation}%
$\varphi (s^{2})$ is the (constrained) instanton profile and $J_{2}(z)$ is
the Bessel function. The explicit form of the Fourier transform of the pure
instanton solution is well known (in the singular gauge)
\begin{equation}
\widetilde{\varphi }^{I}(p^{2})=i\frac{\left( 4\pi \right) ^{2}}{p^{4}}\left[ 1-
\frac{(\rho p)^{2}}{2}K_{2}\left( \rho p\right) \right] ,\ \ \
\widetilde{\varphi }^{I}(p^{2})=\left\{
\begin{array}{c}
\frac{i\left( 2\pi \right) ^{2}\rho ^{2}}{p^{2}},\qquad p^{2}\rightarrow 0,
\\
\frac{i\left( 4\pi \right) ^{2}}{p^{4}},\qquad p^{2}\rightarrow \infty .
\end{array}%
\right.
\label{Phi(p)_SingI_As}\end{equation}
The constraint solution saves its form at short
distances, but changes it at large ones:
\begin{equation}
\widetilde{\varphi }^{CI}(p^{2})=\left\{
\begin{array}{c}
\frac{i\pi ^{2}}{4}R^{4}I_{CI},\qquad p^{2}\rightarrow 0, \\
\frac{i\left( 4\pi \right) ^{2}}{p^{4}},\qquad p^{2}\rightarrow \infty ,%
\end{array}%
\right.  \label{Phi(p)_SingCI_As}
\end{equation}
where the constant $I_{CI}$ is given by
\[
I_{CI}=\int_{0}^{\infty }du\quad u^{2}\varphi (u).
\]%
Now, the Fourier transform of the single instanton contribution to the gluon
propagator (in the Landau gauge) becomes
\begin{eqnarray}
G_{\mu \nu }^{ab}(p) &\equiv &\int d^{4}xe^{ipx}\left\langle 0\left\vert
A_{\mu }^{a,I}(x)A_{\nu }^{b,I}(0)\right\vert 0\right\rangle _{I}=\delta
^{ab}\left( \delta ^{\mu \nu }-\frac{p^{\mu }p^{\nu }}{p^{2}}\right)
G(p),  \label{GProp_I} \\
G(p) &=&-\frac{4}{N_{c}^{2}-1}p^{2}
\int dn(\rho )\rho^{4}\widetilde{\varphi }^{2}( p^2),
\label{GProp_II}
\end{eqnarray}%
where the effective instanton density takes into account averaging over the
instanton size distribution. Thus, we see again that the gluon propagator and
gluon field strength correlator are quite different functions, and the
relation between them valid in the abelian gauge model is destroyed in the non-abelian
case.

From (\ref{Phi(p)_SingI_As}) and (\ref{Phi(p)_SingCI_As})
it is easy to deduce the asymptotics of the instanton
part of the gluon propagator
\begin{equation}
G^{I}(p^{2})=\left\{
\begin{array}{c}
\frac{\left( 2\pi \right) ^{4}n_{c}\rho ^{4}}{N_{c}^{2}-1}\frac{1}{p^{2}}%
,\qquad p^{2}\rightarrow 0 \\
\frac{\left( 4\pi \right) ^{4}n_{c}\rho ^{4}}{N_{c}^{2}-1}\frac{1}{p^{6}}%
,\qquad p^{2}\rightarrow \infty%
\end{array}%
\right. ,\quad G^{CI}(p^{2})=\left\{
\begin{array}{c}
\frac{\pi ^{4}n_{c}R^{4}}{16(N_{c}^{2}-1)}I_{CI}^{2}p^{2},\qquad
p^{2}\rightarrow 0 \\
\frac{\left( 4\pi \right) ^{4}n_{c}\rho ^{4}}{N_{c}^{2}-1}\frac{1}{p^{6}}%
,\qquad p^{2}\rightarrow \infty%
\end{array}%
\right.  \label{GProp_As}
\end{equation}

Calculating (in very similar way as in the Landshoff-Nachtmann model) at
large energy, $s$, the invariant $\mathcal{T-}$matrix element of the
quark-quark scattering exchanging by two gluons we get
\begin{equation}
\left\langle q(p_{3})q(p_{4})\left\vert \mathcal{T}\right\vert
q(p_{1})q(p_{2})\right\rangle \left.
\begin{array}{c}
\rightarrow \\
s\rightarrow \infty%
\end{array}%
\right. iI(t)\quad \overline{u}(p_{3})\gamma ^{\mu }u(p_{1})\overline{u}%
(p_{4})\gamma ^{\mu }u(p_{2}),
\label{Tmatr}\end{equation}
with%
\begin{equation}
I(t)=\frac{1}{2}\int \frac{d\overrightarrow{k_{\perp }}}{(2\pi )^{2}}G\left[
\left( \overrightarrow{k_{\perp }}+\frac{1}{2}\overrightarrow{q_{\perp }}%
\right) ^{2}\right] G\left[ \left( \overrightarrow{k_{\perp }}-\frac{1}{2}%
\overrightarrow{q_{\perp }}\right) ^{2}\right] ,
\label{It}\end{equation}
where $G(p^{2})$ is defined in (\ref{GProp_II}) with
$p^2\to \overrightarrow{p_{\perp }}^2$. Except numerical coefficient, this expression
is in agreement with the Nachtmann-Landshoff formula. This agreement is due to
specific features of the instanton induced interaction.

It is clear from the
infrared behaviour of the instanton induced propagator (\ref{GProp_As}) that
$I(0)$ (\ref{It}) is infinite for the pure instanton solution (\ref{Phi(p)_SingI_As}), but
it is finite for the constraint instanton solution. This fact also noted recently
in \cite{Shuryak:2000df} was one of the
arguments to construct constraint instanton that modifies the profile of the
instanton at large distances.

From (\ref{Tmatr}) and the optical theorem
it follows that the spin averaged total quark-quark cross section is constant
at large energy:
\begin{equation}
\sigma _{qq}\sim (n_{c}\rho _{c}^{4})R^{2}.
\label{Sconst}\end{equation}
These results have been recently generalized in \cite{Shuryak:2000df},
where the growing part of the total cross section was also found
\begin{equation}
\sigma _{qq}\sim (n_{c}\rho _{c}^{4})\Delta (t)\ln s.
\label{Sgrow}\end{equation}

As was discussed in details in \cite{Landshoff:1986yj} already this simple model of the
Pomeron explain many properties of diffractive scattering: the effective vector-like
exchange (\ref{Tmatr}), the additive quark rule and the main features of the total
cross section (\ref{Sconst}), (\ref{Sgrow}).

\section{Instanton Corrections to Quark Form Factor at Large Momentum
Transfer}

One of the important questions in the description of hadronic processes is
the behaviour of the vertex functions (form factors) at various energy
domains. The present section is devoted to the analysis of the instanton
induced corrections to the asymptotic behavior of the color singlet quark
form factor at the large momentum transfer within the framework of the
instanton liquid model of QCD vacuum.

The color singlet quark form factor is determined via the amplitude of
elastic scattering of a quark in electromagnetic field:
\begin{equation}
\mathcal{M}_\mu =F_q[(p_1-p_2)^2] \bar u(p_1) \gamma_\mu \  v(p_2) \ ,
\end{equation}
where $p_1$ and $p_2$ are initial and final momenta of quark of mass $m$.
The kinematics of the process is described (in Minkowski space-time) in
terms of the scattering angle $\chi$:
\begin{equation}
\cosh \chi = {\frac{(p_1 p_2) }{m^2}} = 1 + {\frac{Q^2 }{2 m^2}} \ \ , \ \
Q^2 = - (p_2 - p_1)^2 \ \ , \ \ p_1^2=p_2^2 =m^2 \ .  \label{m1}
\end{equation}

The consistent RG analysis of the total form factor $F_{q}(Q^{2})$ results
in the conclusion that its leading large-$Q^{2}$ behaviour including all the
logarithmic corrections is controlled by the universal cusp anomalous
dimension and can be expressed in the following form \cite{Korchemsky:pn}:
\begin{equation}
F_{q}(Q^{2})=\exp \left[ -\int_{\lambda ^{2}}^{Q^{2}}\!{\frac{d\xi }{2\xi }}
\ \left( \ln {\frac{Q^{2}}{\xi }}\ \Gamma _{cusp}(\alpha _{s}(\xi ))
-{\frac{d\ \ln W_{np}(\xi )}{d\ \ln \xi }}\right) \right]
\ ,  \label{1}
\end{equation}
where $\Gamma _{cusp}(\alpha _{s})$ is the cusp anomalous dimension%
\begin{equation}
\Gamma _{cusp}(\alpha _{s})=\frac{\alpha _{s}(\mu )}{\pi }C_{F}+(\mathrm{%
Higher\quad loops).}
\end{equation}

At high energy the quark receives the eikonal phase during the scattering
process off soft gluons. This effect is expressed in terms of the vacuum
average of the gauge invariant path ordered Wilson integral
%\cite{MMP}
\begin{equation}
W_{np}(C_{\chi })={\frac{1}{N_{c}}}Tr\langle 0|\mathcal{P}\exp \left(
ig\int_{C_{\chi }}\!dx_{\mu }\hat{A}_{\mu }(x)\right) |0\rangle \ ,
\label{1a}\end{equation}
In Eq. (\ref{1a}) the integration goes along the closed contour $C_{\chi }$
with cusp.

In the single instanton sector the Wilson integral (\ref{1a}) is in the form:
\begin{equation}
w_{I}(C_\chi)={\frac{1}{N_{c}}}\langle 0|Tr\exp \left( i\sigma ^{a}\phi
^{a}\right) |0\rangle \ ,  \label{wI1}
\end{equation}
\begin{equation}
\phi ^{a}={R}^{ab}{\eta ^{\pm }}_{\mu \nu }^{b}\int_{C_{\gamma }}\!dx_{\mu
}\ (x-z_{0})_{\nu }\varphi (x-z_{0};\rho )\ .  \label{iin}
\end{equation}%
We omit the path ordering operator $\mathcal{P}$ in (\ref{wI1}) because the
instanton field (\ref{Agen,F}) is a hedgehog in color space.
Note, that for the instanton calculations, it is necessary to map the scattering
angle to the Euclidean world by the analytical continuation
$\chi\rightarrow i\gamma \ $, and perform the inverse transition in the final
expressions in order to restore the $Q^{2}$-dependence.
Therefore,
we find that the instanton contribution to the quark form factor \cite{Dorokhov:2002qf}:
\be
F_q(Q^2) = \exp{\left[- {2C_F \over \beta_0} \ln {Q^2 \over \Lambda^2} \left(
\ln{\ln(Q^2/\Lambda^2) \over \ln(\lambda^2/Q_0^2)} \right) + \ln {Q^2 \over Q_0^2}
\left({2C_F \over \beta_0} - B_I \right)  \right]} \ .  \label{eq:final} \ee
results in a finite renormalization of the
next-to-leading logarithmic perturbative term.

In the leading order in instanton field the nonperturbative
contribution can be expressed in the form (here we use the
exponentiation of the single-instanton result in a dilute instanton ensemble
\cite{Dorokhov:2002qf}):
\begin{equation}
{\frac{d\ \ln W_{np}^{I}(Q^{2})}{d\ \ln Q^{2}}}={\frac{\pi ^{2}}{2}}\int
\!dn(\rho )\ \rho ^{4}\ln {\left( \rho ^{2}\lambda ^{2}\right) }\equiv B_I\ .
\label{II1}
\end{equation}

In order to estimate the magnitude of the instanton induced effect we
consider the standard distribution function %\cite{tH}
supplied with the
exponential suppressing factor, what has been suggested in \cite{Shuryak:1999fe} (and
discussed in \cite{Dorokhov:1999ig} in the framework of constrained instanton model)
in order to describe the lattice data \cite{Hasenfratz:1998qk}:
\begin{equation}
dn(\rho )={\frac{d\rho }{\rho ^{5}}}\ C_{N_{c}}\left( \frac{2\pi }{\alpha
_{s}(\mu _{r})}\right) ^{2N_{c}}\exp \left( -{\frac{2\pi }{\alpha _{s}(\mu
_{r})}}\right) \left( \rho \mu _{r}\right) ^{b}\exp \left( -2\pi \sigma \rho
^{2}\right) \ ,  \label{dist1}
\end{equation}%
where the constant $C_{N_{c}=3}\approx0.0015$,
$\sigma $ is the string tension.
Given the distribution function (\ref{dist1}) the parameters of
the instanton liquid: the mean instanton size $\bar{\rho}$ and the instanton
density $\bar{n}$ will read:
\begin{equation}
\bar{\rho}={\frac{\Gamma (b/2-3/2)}{\Gamma (b/2-2)}}{\frac{1}{\sqrt{2\pi
\sigma }}}\ ,\quad \bar{n}={\frac{C_{N_{c}}\Gamma (b/2-2)}{2}}\left( \frac{%
2\pi }{\alpha _{s}(\bar{\rho})}\right) ^{2N_{c}}\left( {\frac{\Lambda _{QCD}%
}{\sqrt{2\pi \sigma }}}\right) ^{b}(2\pi \sigma )^{2}\ .  \label{nbar}
\end{equation}

By using the one loop approximation for the running coupling constant,$%
\alpha _{s}(\mu )={\frac{4\pi }{\beta _{0}\ln \ {\mu ^{2}/\Lambda ^{2}}},}$ $%
\beta _{0}={\frac{11N_{c}-2n_{f}}{3},}$ we find the instanton contribution (%
\ref{II1}) in the form:
\begin{equation}
{\frac{d\ \ln W_{np}^{I}(Q^{2})}{d\ \ln Q^{2}}}=-{\frac{K}{\beta _{0}}}\pi
^{2}\bar{n}{\bar{\rho}}^{4}\ln {\frac{2\pi \sigma }{\lambda ^{2}}}\ ,\quad
K\approx 0.74\ .  \label{pow1}
\end{equation}%

In Eq. (\ref{nbar}) we choose the normalization scale $\mu
_{r}$ of order of the instanton inverse mean size $\bar{\rho}^{-1}$.
The packing fraction $\pi ^{2}\bar{n}{\bar{\rho}}^{4}$ characterizes
diluteness of the instanton liquid and within the conventional picture its
value is estimated to be $\approx 0.12\ $,
if one takes the model parameters as
%(see \cite{REV}):
\begin{equation}
{\bar n}\approx 1fm^{-4},\ \ {\bar \rho} \approx 1/3fm\ ,\ \ \sigma =(0.44\ GeV)^{2}.
\label{param}
\end{equation}%
The leading contribution to the quark form factor at asymptotically large $%
Q^{2}$ is provided by the (perturbative) evolution governed by the cusp
anomalous dimension
Thus, the instantons yield the sub-leading
effects to the large-$Q^{2}$ behaviour accompanied by a numerically small
($B_I\sim 0.01$) factor compared to perturbative one ($\approx 0.2$). Therefore,
the effect of the instantons results in a finite renormalization of the next-to-leading logarithmic
perturbative term in the exponentiated expression.

\section{Pion Transition Form Factor}

Let us consider the pion
 form factor for the transition process $\gamma ^{\star
}\gamma^{\star}\rightarrow \pi ^{0}$ at space-like values of photon momenta.
The interest to the pion transition form factor has recently revived due to
its measuring by CLEO collaboration \cite{Gronberg:1997fj} at large
virtuality of one of the photons.  Theoretically, the pion form factor $M_{\pi
^{0}}(q_{1}^{2},q_{2}^{2})$ for the transition process $\gamma ^{\star
}(q_{1})\gamma ^{\star }(q_{2})\rightarrow \pi ^{0}(p)$, where $q_{1}$ and $%
q_{2}$ are photon momenta, is related to fundamental properties of QCD
dynamics at low and high energies. At zero photon virtualities the value of
the form factor and its slope (radius) is estimated within the chiral
perturbative theory. In the opposite limit of large photon virtualities the
leading momentum power dependence \cite{Lepage:1979zb} of the form factor
supplemented by small radiative \cite{delAguila:1981nk} and power corrections%
 \cite{Chernyak:1982is} is dictated by perturbative QCD (pQCD).

In this section, we discuss the approach \cite{Anikin:rq,Dorokhov:2001wx} that allow us to
match these extremes and describe the intermediate energy region. This
approach describes quark-meson dynamics within the effective model, where
the quark-quark interaction induced by instanton exchange leads to
spontaneous breaking of the chiral symmetry. It dynamically generates the
momentum dependent quark mass $M(k^{2})$ that may be related to the
quark nonlocal condensate \cite{Dorokhov:1997iv}. Specifically, we find
\cite{Dorokhov:2002iu,Dorokhov:2002rc} the pion transition form factor in wide kinematical region
up to moderately large photon virtualities and extract from its asymptotics
the pion distribution amplitudes (DAs) at normalization scale typical for
hadrons.

The invariant amplitude for the process $\gamma ^{\ast }\gamma
^{\ast}\rightarrow \pi ^{0}$ is given by
\[
A\left( \gamma ^{\ast }\left( q_{1},\epsilon _{1}\right) \gamma ^{\ast
}\left( q_{2},\epsilon _{2}\right) \rightarrow \pi ^{0}\left( p\right)
\right) =-ie^{2}\varepsilon _{\mu \nu \rho \sigma }\epsilon _{1}^{\mu
}\epsilon _{2}^{\nu }q_{1}^{\rho }q_{2}^{\sigma }M_{\pi ^{0}}\left(
q_{1}^{2},q_{2}^{2}\right) ,
\]%
where $\epsilon _{i}^{\mu }(i=1,2)$ are the photon polarization vectors.
Consider first the low energy region. With both photons real $\left(
q_{i}^{2}=0\right) $ one finds the result
\begin{equation}
M_{\pi ^{0}}\left( 0,0\right) =\frac{N_{c}}{6\pi ^{2}f_{\pi }}%
\int_{0}^{\infty }du\frac{uM(u)\left[ M(u)-2uM^{\prime }(u)\right] }{%
D^{3}(u) }=\frac{1}{4\pi ^{2}f_{\pi }},  \label{ChAn}
\end{equation}
where $D(u)=u+M^{2}(u)$ and $M^{\prime }(u)=\frac{d}{du}M(u),$ consistent
with the chiral anomaly and independent of the shape of $M(k^{2}).$ Below,
for the numerical analysis we choose the dynamical quark mass profile in the
Gaussian form $M_{G}(k^{2})=M_{q}\exp {(-2k^{2}/\Lambda ^{2})},$ where we
take $M_{q}=350$ MeV and fix $\Lambda =1.29$ GeV from the pion weak decay
constant, $f_{\pi }=92.4$ \textrm{MeV}. We also consider the shape given by
the quark zero modes (z.m.) in the instanton field: $%
M_{I}(k^{2})=M_{q}Z^{2}(k\rho )$, where $Z(k\rho )=2z\left[
I_{0}(z)K_{1}(z)-I_{1}(z)K_{0}(z)-I_{1}(z)K_{1}(z)/z\right] _{z=k\rho /2}$,
with $\rho =1.7$ GeV$^{-1}$ being the inverse mean instanton radius and $%
M_{q}=345$ MeV. The mean square radius of the pion for the transition $%
\gamma ^{\ast }\pi ^{0}\rightarrow \gamma $ is found to be $r_{\pi \gamma
}^{2}\approx 1/(2\pi^2f_\pi^2)$ and is almost independent on the form of $%
M(k^{2}).$

At large photon virtualities $Q^{2}=-(q_{1}^{2}+q_{2}^{2})$ the model
calculations reproduce the pQCD factorization result ($%
\omega=(q_{1}^{2}-q_{2}^{2})/(q_{1}^{2}+q_{2}^{2})$)
\begin{equation}
\left. M_{\pi ^{0}}(q_{1}^{2},q_{2}^{2})\right\vert _{Q^{2}\rightarrow
\infty }=J^{(2)}\left( \omega \right) \frac{1}{Q^{2}}+J^{(4)}\left( \omega
\right) \frac{1}{Q^{4}}+O(\frac{\alpha _{s}}{\pi })+O(\frac{1}{Q^{6}}).
\label{AmplAsympt}
\end{equation}%
The leading and next-to-leading order asymptotic coefficients
\begin{eqnarray}
J^{(2)}\left( \omega \right) &=&\frac{4}{3}f_{\pi }\int_{0}^{1}dx\frac{%
\varphi _{\pi }^{(2)}(x)}{1-\omega ^{2}(2x-1)^{2}},\ \ \   \nonumber \\
J^{(4)}\left( \omega \right) &=&\frac{4}{3}f_{\pi }\Delta ^{2}\int_{0}^{1}dx%
\frac{1+\omega ^{2}(2x-1)^{2}]}{[1-\omega ^{2}(2x-1)^{2}]^{2}}\varphi _{\pi
}^{(4)}(x)  \label{J}
\end{eqnarray}%
are expressed in terms of the light-cone pion distribution amplitudes (DA), $%
\varphi _{\pi }(x)$, that are predicted \cite{Dorokhov:2002iu} by the model at the low
normalization scale $\mu ^{2}\sim \Lambda ^{2}\sim \rho^{-2}$
%(Fig. 1, for Gaussian $M(k^2)$)
\begin{equation}
\varphi _{\pi }^{(2)}(x)=\frac{N_{c}}{4\pi ^{2}f_{\pi }^{2}}\int_{-\infty
}^{\infty }\frac{d\lambda }{2\pi }\int_{0}^{\infty }du \frac{F(u_+,u_-)}{%
D\left( u_+\right) D\left( u_-\right) } \left[ xM\left( u_+\right)
+\left(x\leftrightarrow \overline{x}\right) \right] ,  \label{WF_VF2}
\end{equation}
\begin{equation}
\varphi _{\pi }^{(4)}(x)=\frac{1}{\Delta ^{2}}\frac{N_{c}}{4\pi
^{2}f_{\pi}^{2}} \int_{-\infty }^{\infty }\frac{d\lambda }{2\pi }%
\int_{0}^{\infty }du \cdot\frac{uF(u_+,u_-)}{D\left( u_-\right) D\left(
u_-\right) }\left[ \overline{x}M\left(u_+\right) +\left( x\leftrightarrow
\overline{x}\right) \right] ,  \label{WF_VF4}
\end{equation}
where $u_+=u+i\lambda\overline{x}$, $u_-=u-i\lambda x$, $F\left( u,v\right) =%
\sqrt{M\left( u\right) M\left( v\right) },$ and $\overline{x}=1-x$. The
parameter $\Delta ^{2}$ characterizing the scale of the power corrections in
the hard exclusive processes is
\begin{equation}
\Delta ^{2}=\frac{N_{c}}{4\pi ^{2}f_{\pi }^{2}}\int_{0}^{\infty }du\frac{%
u^{2}M(u)(M(u)+\frac{1}{3}uM^{\prime }(u))}{D^{2}(u)},  \label{PowCorr}
\end{equation}
Its value is predicted $\Delta^{2}= 2.41(2.74)\pi^2f_\pi^2$ for Gaussian
(zero mode) shape of $M(k^2)$, correspondingly.
%As it is clear from Fig. 1,
The leading order pion DA, $\varphi _{\pi }^{(2)}(x)$, is close to the
asymptotic form that is in agreement with the results obtained previously in%
 \cite{Mikhailiov:1989mk,Kroll:1996jx}. In the leading order the similar
results within the instanton model have been derived earlier in \cite%
{Esaibegian:1989uj}.

The asymptotic coefficients $J(\omega )$ may be written in the form \cite{Dorokhov:2002iu}
\[
J^{(2)}\left( \omega \right) =-\frac{1}{\pi^2f_\pi}\int_{0}^{\infty
}duu\int_{0}^{\infty }dv\cdot
\]
\begin{equation}
\cdot\left\{ \frac{M^{1/2}\left( z_{-}\right) }{D(z_{-})}\frac{\partial }{%
\partial z_{+}}\left( \frac{M^{3/2}\left( z_{+}\right) }{D(z_{+})}\right)
+\left( z_{-}\longleftrightarrow z_{+}\right) \right\} ,  \label{J2}
\end{equation}
\[
J^{(4)}\left( \omega \right) =\frac{2}{\pi^2f_\pi}\int_{0}^{\infty
}du\int_{0}^{\infty }dvv\cdot
\]
\begin{equation}
\cdot\left\{ \frac{M^{1/2}\left( z_{-}\right) }{D(z_{-})}\left[ \frac{%
M^{3/2}\left( z_{+}\right) }{D(z_{+})}+u\frac{\partial }{\partial z_{+}}%
\left( \frac{M^{3/2}\left( z_{+}\right) }{D(z_{+})}\right) \right] +\left(
z_{-}\longleftrightarrow z_{+}\right) \right\} ,  \label{J4}
\end{equation}%
where $z_{\pm }=u+v(1\pm \omega )$. With the model parameters given above we
find for the process $\gamma \gamma ^{\ast }\rightarrow \pi ^{0}$ the values
$J^{(2)}\left( \omega =1\right) =1.83(2.13)f_\pi$ consistent with the CLEO
fit $J_{\exp }^{(2)}\left( 1\right) =(1.74\pm 0.32)f_\pi$ and the power
correction $J^{(4)}\left( 1\right) /J^{(2)}\left( 1\right)
=2.97(3.62)\pi^{2}f_\pi^2$.
The model form factors \cite{Dorokhov:2002iu}
take into
account the perturbative $\alpha_{s}(Q^{2})-$ corrections \cite%
{delAguila:1981nk} to the leading twist-2 term with the running coupling
that has zero at zero momentum. The perturbative corrections to the twist-4
contribution and the power corrections generated by the twist-3 pion DAs are
expected to be inessential.

Thus, within the covariant nonlocal model describing the quark-pion dynamics
we obtain the $\pi \gamma ^{\ast }\gamma ^{\ast }$ transition form factor in
the region up to moderately high momentum transfer squared, where the rapid
power-like asymptotics takes place. At larger virtualities the pQCD
evolution of the DA slowly goes to the asymptotic limits. From the
comparison of the kinematical dependence of the asymptotic coefficients of
the transition pion form factor, as it is given by pQCD and the
nonperturbative model, the relations between the pion DAs and the dynamical
quark mass and quark-pion vertex are derived. The leading and next-to-leading
order power asymptotics of the form factor
and the relation between the light-cone pion distribution amplitudes of
twists 2 and 4 and the dynamically generated quark mass are found.

%\section{Discussion}

{\small The author thanks the organizers for creation of very fruitful atmosphere
during the school.
This work is supported by RFBR Grants 01-02-16431, 02-02-81023 and 02-02-16194
and by INTAS Grant 2000-366.}

%\lastevenpage
\end{document}